\documentclass[12pt]{iopart}
\usepackage{iopams}
\expandafter\let\csname equation*\endcsname\relax

\expandafter\let\csname endequation*\endcsname\relax

\usepackage{mathrsfs,amsmath}
\usepackage[scr=boondox]{mathalfa}
\usepackage{cite,comment}
\usepackage{graphicx, float}
\usepackage{dcolumn}
\usepackage{bm}
\usepackage{makecell}
\usepackage{epstopdf}
\usepackage{doi,hyperref,url}
\usepackage[usenames, dvipsnames]{color}

\usepackage[caption=false]{subfig}

\begin{document}

\title[Effect of viscosity and thermal conductivity on the radial oscillation]{Effect of viscosity and thermal conductivity on the radial oscillation and relaxation of relativistic stars}

\author{D\'{a}niel Barta${}^{1}{}^{2}$}
\address{${}^{1}$ Wigner RCP, Konkoly-Thege Mikl\'{o}s \'{u}t 29-33., H-1121 Budapest, Hungary}
\address{${}^{2}$ OzGrav: The ARC Centre of Excellence for Gravitational-wave Discovery, Clayton, VIC 3800, Australia}
\ead{barta.daniel@wigner.mta.hu}
\vspace{10pt}
\begin{indented}
\item[]August 1, 2019
\end{indented}

\begin{abstract}
In this paper we present a generic formulation of the linearized dynamical equations governing small adiabatic radial oscillations of relativistic stars. The dynamical equations are derived by taking into consideration those effects of viscosity and thermal conductivity of neutron-star matter which directly determine the minimum period of observable pulsars. A variational principle is applied to determine a discrete set of eigenfunctions with complex eigenvalues. The real and imaginary parts of eigenvalues represent the squared natural frequencies and relaxation time of radial oscillations of non-rotating neutron stars, respectively. We provide a suitable framework which may be supplemented with various potential species of cold-nuclear-matter models to compute the spectra of the normalized eigenfrequencies with a certain numerical precision. In the last section, we provide a qualitative estimation of the rate at which viscosity and thermal conductivity drain the kinetic energy of radial oscillation mode in reasonably uniform neutron stars, without relying on explicit numerical computations.
\end{abstract}

%
%
%
%
%

\section{\label{sec:intro}Introduction}
\subsection{Astrophysical motivation and relation to earlier works}
Neutron star oscillations may impact on a range of observations, involving in particular radio and X-ray timing and gravitational waves. The possibility that radial oscillations of neutron stars give rise to oscillations observed within radio subpulses of pulsars was proposed by V. Boriakoff \cite{Boriakoff1976} in 1976. X-ray and $\gamma$-ray burst events have been generally associated with neutron stars by many authors e.g. \cite{Ramaty1981}. Periodicities have been observed in X-ray bursts, which has raised considerable interest in radial \cite{Glass1983a} and other types \cite{Glass1983b} of neutron-star oscillations since the early 1980s. Bursters exhibit periodic and rapid increases in luminosity (typically a factor of 10 or greater) when enormous amount of energy deposited in oscillation modes is released in a short period of time through heat outflow via neutrino emission. \cite{Chirenti2017} Physically, some mass from the stellar interior is drawn toward the surface where the hydrogen fuses to helium which accumulates until it fuses in a burst, producing X-rays. After the subsequent emission of thermal photon from the surface, the crust thermally relaxes toward equilibrium with the core. X-ray observations from the recently lunched NICER mission \cite{Gendreau2012} and from the upcoming LOFT mission \cite{Watts2016} will yield the mass and radius of a few stars up to $\sim 5\%$ precision. The observations of binary NS and BHNS coalescence events by gravitational-wave detectors such as the transient signal GW170817 will also dramatically improve our understanding of ultra-dense matter in neutron stars. Further interest in the study has been stimulated by \cite{Haensel1991}, where $\gamma$-ray bursts were assumed to originate as results of collisions between strange stars at cosmological distances. Even though radial oscillations of strange stars are expected to be damped rapidly \cite{Haensel1989}, such strange stars during their short time-scales are likely to be promising targets for multi-messenger observations. For most stellar models, the periods (typically ranging from 0.2 to about 0.9 milliseconds) depend on the stellar model and its central density \cite{Detweiler1983}, while the relaxation time is in the range of  $0.1 - 0.3$ seconds. Therefore the study of oscillation spectra and relaxation times of neutron stars could be very valuable as their dynamical behaviour may eventually be deduced from observations.

Theoretical interest in the dynamical stability of relativistic stars has arisen since 1964 from the seminal works of S. Chandrasekhar, R.F. Tooper and J.M. Bardeen \cite{Chandrasekhar,Chandrasekhar2,Tooper1964,Tooper1965,Bardeen1965} and a general stability criterion was formalized in the 1970s by J. L. Friedman and B. F. Schutz \cite{Friedman1975}. The stability of spherically symmetric stars under radial adiabatic perturbations has been extensively studied and reported in the literature (e.g \cite{Glass1983a,Glass1983b,Harrison1965,Chanmugam1977,Vath1992}). Induced by the small radial perturbations, dynamical instability will intervene by radial oscillation before the star contracts. Several techniques for obtaining spectra of oscillation modes have been developed for various stellar equilibrium models, mostly with zero-temperature equation-of-state (EOS). We have applied Chandrasekhar's linear varational method \cite{Chandrasekhar,Chandrasekhar2} to formulate the variational principle which forms the basis for determining the characteristic eigenfrequencies and relaxation times of radial oscillations.

Concurrently, the conversion of kinetic energy into heat and effects of viscosity on stellar pulsations in general has been addressed first by \cite{Kopal1964,Higgins1968,Mihalas1983}. Properties of transport coefficients (bulk viscosity, shear viscosity, thermal conductivity) in neutron stars have been studied more in detail by a number of recent of works \cite{Shternin2008,Shternin2013,Shternin2017,Tolos2016}. The density and temperature dependence of shear viscosity and of bulk viscosity in the crust and in the core, respectively, have been described for different EOS of neutron stars by \cite{Shternin2008}. Thermal conductivity and shear viscosity of nuclear matter arising from nucleon-nucleon interaction in non-superfluid neutron-star cores were considered by \cite{Shternin2013}, whereas those arising do from the collisions among phonons in superfluid neutron stars were considered by \cite{Tolos2016}. An extension of \cite{Shternin2013} for different nucleon-nucleon potentials and different three-body forces in \cite{Shternin2017} found that the nucleon contribution dominates the thermal conductivity, but the shear viscosity is dominated by leptons. The most up-to-date review of the transport properties and the underlying reaction rates of dense hadronic and quark matter in the crust and the core of neutron stars is found in \cite{Schmitt2018}.

\subsection{Objectives and content}
The present paper, together with its accompanying paper \cite{Barta2019}, is part of a more extensive body of work investigating the radial oscillations of neutron stars affected by the viscosity and thermal conductivity of cold nuclear matter that brings about damping of stellar oscillations and directly determine a minimum period of observable pulsars. The separation of the analytical and numerical aspects of study into two distinct papers was felt desirable, partly in order to not to divert attention from the details of each respective aspect, partly for the inherently different approaches involved.

Here, in addition to providing a comprehensive review of the basic theory of relativistic stellar pulsations, we present an analytical formulation of the dynamical equations that governs the radial mode of linear adiabatic stellar oscillations through a perturbation scheme. We prove that, similarly to the non-dissipative case, the pulsation equation expressed by a set of effective variables which involve dissipative terms, can be recast in a self-adjoint form. In contrast to the common non-dissipative case, the associated Sturm--Liouville eigenvalue problem (SLEVP) is generalized for a discrete set of eigenfunctions with complex eigenvalues which correspond to the squared frequencies of the oscillation modes and the imaginary part corresponds to the damped solution. However, the main novelty of this approach is the ability to directly relate the damping ratio of oscillations to the expressions $\mathcal{S}_{1}$ and $\mathcal{S}_{2}$, which stem from the viscous and heat-conductive contributions to the stress--energy tensor, without relying on explicit numerical computations. The scale of relaxation time, directly related to the damping ratio, is identical to approximate solution for the time-scale of energy dissipation given by \cite{Cutler} in an alternative way. The usefulness of our analytical approximation method is evidently restricted to providing qualitative and ``order-of-magnitude'' information about the dissipative time-scales in \eqref{dissipative-time-scale2} rather than a precise one. 

Conversely, the numerical solution of the eigenvalue problem is considered in the accompanying paper \cite{Barta2019}. The SLEVP for the radial oscillation modes of stars is converted to a system of finite difference equations where we implement a second-order accurate differencing scheme so the resulting system of finite difference equations emerges as a tridiagonal matrix eigenvalue problem. In a manner similar to Kokkotas and Ruoff (2001) \cite{Kokkotas} approach, we compute the four lowest-frequency radial-oscillation modes of neutron stars constructed from various potential EOS of cold-nuclear-matter considered by \"{O}zel and Freire (2016) \cite{Ozel2016}. The algorithm yields zero-frequency modes at the maxima and minima of the mass curves while the equilibrium adiabatic index characterizes the stiffness of the EOS at a given density. Finally, we evaluate the rate at which viscosity and thermal conductivity drain energy from the radial oscillation mode. \\

The present paper is organized as follows: Mostly based on \cite[p.~285--289]{Rezzola}, sec. \ref{sec:nonperfect} gives a brief relativistic description of perfect fluids and of non-perfect fluids (viscous and heat-conductive) in Eckart frame (or 'particle frame'). The Einstein field equations and the generalized Tolman--Oppenheimer--Volkoff equation are expressed in sec. \ref{sec:fieldeqs} through effective variables that incorporate time-dependent dissipative contributions of the neutron-star matter. In sec. \ref{sec:oscillations}, we present a variational method to formulate a lowest-order asymptotic approximation of infinitesimal adiabatic radial oscillations. We adopt a linear perturbation theory from \cite{Chandrasekhar} for the Lagrangian displacement of fluid elements from the equilibrium state where the adiabatic perturbations are considered to be purely radial. Stellar equilibrium models are described in sec. \ref{sec:gamma} through a local adiabatic index which can be regarded as constant near the center, but in general, varies, depending on the dynamical regime. Sec. \ref{sec-pulsation} discusses the second-order linear ordinary differential equation of radial pulsations \eqref{pulsation-eq2}, derived from the perturbation equations. Sec. \ref{sec:eigenfrequency} deals with this pulsation equation that imposes a regular SLEVP for natural frequencies of oscillation with separated boundary conditions \eqref{boundary-condition2} induced by evaluating and setting $\chi$ to zero at the boundary. The characteristic time-scale for the relaxation of radial oscillation mode due to viscous dissipation is discussed in sec. \ref{sec:relaxation-time}, based on \cite{Cutler}. Lastly, the principal results established in this paper are summarized in sec. \ref{sec:conclusion}.

\section{\label{sec:nonperfect}Stress--energy tensor of perfect and of non-perfect fluids}
We may first consider a finite spherical distribution of fluids bounded by a sperical surface $\Sigma$ and held together by its own gravitational attraction. The line element for such a system is given in Schwarzschild-like coordinates by
\begin{equation} \label{line-element}
	ds^{2} = e^{\nu}dt^{2} - e^{\lambda}dr^{2} - r^{2}(d\vartheta^{2} + \sin^{2}\vartheta d\varphi^{2})
\end{equation}
with the metric potentials $\nu = \nu(t,r)$ and $\lambda = \lambda(t,r)$ being dependent on the temporal and radial coordinates $(t,\, r)$. The general-relativistic hydrodynamic equations for a generic fluid involve the equations of motion that are given by the conservation of rest mass and by the conservation of energy and momentum:
\begin{equation} \label{conservation-of-mass}
	\nabla_{\mu}J^{\mu} = 0,
\end{equation}
\begin{equation} \label{conservation-of-energy}
	\nabla_{\mu}T^{\mu\nu} = 0,
\end{equation}
respectively, and the second law of thermodynamics
\begin{equation} \label{entropy-principle}
	\nabla_{\mu}S^{\mu} \geq 0
\end{equation}
that are relativistically consistent. Although the perfect-fluid approximation disregards scenarios when dissipation and energy fluxes are present, it works well for most fluids under generic conditions. However, it loses its validity when thermodynamic (i.e microscopic) time-scales are comparable to the dynamic (i.e. macroscopic) ones and thus when assumption of local thermodynamic equilibrium breaks down. The requisite extension of perfect-fluid description that accounts for dissipative terms and energy fluxes is non-perfect fluid. In general, one can assume the rest-mass density current and stress--energy tensor as the linear combination of two contributions:
\begin{equation} \label{lin-comb}
	\displaystyle J^{\mu} = J^{\mu}_{\text{PF}} + J^{\mu}_{\text{NPF}}, \quad  T^{\mu\nu} = T^{\mu\nu}_{\text{PF}} + T^{\mu\nu}_{\text{NPF}},
\end{equation}
where the indices "PF" and "NPF" refer to the perfect and non-perfect fluid contributions, respectively.

\subsection{Perfect fluids}
For a system which consists of perfect fluid with total-energy density $\epsilon$, isotropic pressure $p$ and covariant metric elements $g_{\mu\nu}$ corresponding to the antecedent line element \eqref{line-element}, one shall have
\begin{equation} \label{perfect-energy-momentum-tensor}
	\displaystyle J_{\mu}^{\text{PF}} = \rho u_{\mu}, \quad T_{\mu\nu}^{\text{PF}} = (\epsilon + p)u_{\mu}u_{\nu} - pg_{\mu\nu},
\end{equation}
where the spatial components of fluid four-velocity $u_{\mu}$ are zeros. Normalized to $u^{\mu}u_{\mu} = 1$, it becomes
\begin{equation}
	u_{\mu} = (e^{\nu/2},\, 0,\, 0,\, 0).
\end{equation}
By construction, the quantity $\epsilon$ introduced above in eq. \eqref{perfect-energy-momentum-tensor} represents the total-energy density of the fluid, given by
\begin{equation} \label{energy-density}
	\epsilon = \rho(1 + \varepsilon)
\end{equation}
which consists of both the rest-mass density of the fluid $\rho$ and the specific internal-energy density $\varepsilon$, internal-energy density per unit rest mass or which in this case represents the thermal motion of the constituent fluid particles. \cite[p.~98]{Rezzola} Finally, we define the specific enthalpy $h$ as
\begin{equation}
	h = \frac{p + \epsilon}{\rho} = 1 + \varepsilon + \frac{p}{\rho}.
\end{equation}
Now, recognizing that in a non-relativistic regime $\varepsilon \ll c^{2}$ (i.e., the energy density of the fluid is essentially given by the rest-mass density) and $p/\rho \ll c^{2}$ (i.e., the pressure contribution to the energy density is negligible), the Newtonian limit of the specific enthalpy is given by
\begin{equation}
	h = 1 + \varepsilon + \frac{p}{\rho} \to 1.
\end{equation}
Note that there are two natural ways to define four-velocity $u^{\mu}$. One option, given by Eckart, uses a unit timelike vector $\textbf{\textit{u}}_{\text{N}}$ parallel to $\textbf{\textit{J}}$, whereas the other, suggested by Landau, defines a unit timelike vector $\textbf{\textit{u}}_{\text{E}}$ parallel to $\textbf{\textit{T}}\cdot\textbf{\textit{u}}_{\text{E}}$. However, these two vectors are identical for a perfect fluid and parallel to entropy current \textbf{\textit{S}}. The (maximum) entropy principle \eqref{entropy-principle} implies a strict equality for perfect fluids, whose entropy current is then given simply as $S^{\mu} = s\rho u^{\mu}$. However, for relation \eqref{entropy-principle} to be strictly non-zero, the entropy current must have an additional contribution from dissipative parts (first-order theories) with non-zero divergence such that
\begin{equation}
	S^{\mu} = s\rho u^{\mu} + q^{\mu}/T,
\end{equation}
where the temperature $T$, deduced from the first law of thermodynamics is given by
\begin{equation}
	T = \frac{1}{\rho}\left(\frac{\partial \epsilon}{\partial s}\right)_{\rho}
\end{equation}
and
\begin{equation}
	q^{\mu} = (0,q^{i})
\end{equation}  
are the components of the heat-flux four-vector that describes the rate of energy flow per unit area along each spatial coordinate axis within the Eckart frame.  We will adopt the Eckart frame in which the continuity equation retains the same equation \eqref{conservation-of-energy} as for a perfect fluid and there is no dissipative contribution to the the rest-mass density current and to the energy density.

\subsection{Non-perfect fluids}
The dissipative contributions  $T^{\mu\nu}_{\text{NPF}}$ to the stress--energy tensor in eq. \eqref{lin-comb} can be further decomposed into the form
\begin{equation}
	T^{\mu\nu}_{\text{NPF}} = \mathcal{S}^{\mu\nu} + T^{\mu\nu}_{\text{flux}},
\end{equation}
where $\mathcal{S}^{\mu\nu}$ is referred to as viscous stress tensor. An appropriate definition that avoids the possibility of superluminal propagation of speed is given by
\begin{equation}
	\mathcal{S}^{\mu\nu} = \pi^{\mu\nu} + \Pi h^{\mu\nu},
\end{equation}
where $\pi^{\mu\nu}$ is the anisotropic stress tensor, $\Pi = p - p_{\text{eq}}$ is the viscous bulk pressure that measures the deviation of pressure from its equilibrium value, and
\begin{equation} \label{projection}
	h_{\mu\nu} = g_{\mu\nu} - u_{\mu}u_{\nu},
\end{equation}  
is the standard projection tensor onto 3-space normal to flow; all tensors are symmetric. The heat-flow tensor
\begin{equation}
	T^{\mu\nu}_{\text{flux}} = u^{(\mu}q^{\nu)}
\end{equation}  
accounts for the generation of energy fluxes. Altogether, $\mathcal{S}^{\mu\nu},\, \pi^{\mu\nu}$ and $\Pi$ are known as the thermodynamic fluxes and they account for the deviations of the fluid from a perfect fluid. Finally, the full stress--energy tensor will take the form
\begin{equation} \label{full-energy-momentum}
	T^{\mu\nu} = (\epsilon + p)u^{\mu}u^{\nu} + (p+\Pi)h^{\mu\nu} + \pi^{\mu\nu} + u^{(\mu}q^{\nu)}.
\end{equation}

All the different equations for the thermodynamical fluxes $(\pi_{\mu\nu},\, \Pi,\, q_{\mu})$ reduce under the condition \cite{Cutler} that
\begin{enumerate}
	\item the dissipation coefficients are sufficiently small and
	\item the time derivates of perturbed quantities remain sufficiently small
\end{enumerate}
or in other words: provided that $T\nabla_{\mu}S > 0$, i.e. the entropy four-divergence satisfy the condition of being non-negative \cite{Rezzola}, the relations of thermodynamical fluxes to thermodynamic forces $(\Theta,\, h^{\nu}_{\;\mu}\ln T_{,\nu} + u_{\mu;\nu}u^{\nu},\, \sigma_{\mu\nu})$ become linear: 
\begin{equation} \label{constitutive-eqs}
	\begin{array}{l}
		\pi_{\mu\nu} = -2\eta\sigma_{\mu\nu}, \\[10pt]
		\Pi = -\zeta\Theta, \\[10pt]
		q_{\mu} = -\kappa\left(h^{\nu}_{\;\mu}T_{,\nu}-Tu_{\mu;\nu}u^{\nu}\right),
	\end{array}
\end{equation}
where the symmetric, trace-free, spatial shear tensor $\sigma_{\mu\nu}$ is defined by
\begin{equation} \label{shear-tensor}
	\sigma_{\mu\nu} = h^{\alpha}_{\;\mu}h^{\beta}_{\;\nu}u_{(\alpha;\beta)} - \frac{2}{3}\Theta h_{\mu\nu}
\end{equation}
and expansion scalar (or dilatation rate)
\begin{equation} \label{dilatation-rate}
	\Theta = u^{\mu}_{\; ;\mu}
\end{equation}
is associated with the divergence or convergence of the fluid world lines. \cite{Madore} $\eta$ is the shear (also 'common' or 'dynamic') viscosity coefficient that describes the fluid's resistance to gradual shear deformation; $\zeta$ is the  bulk (or 'second') viscosity coefficient that defines the resistance of the medium to gradual uniform compression or expansion; and $\kappa$ is non-negative and accounts for the thermal conductivity, respectively. Collectively, $\kappa,\, \zeta$, and $\eta$ are called transport coefficients or dissipation coefficients, while eqs. \eqref{constitutive-eqs} are referred to as the constitutive equations of classical irreversible thermodynamics, also known as Eckart's theory of relativistic irreversible thermodynamics \cite{Eckart}. Eqs. (\ref{constitutive-eqs}--\ref{dilatation-rate}) give rise to an alternative form of the full stress--energy tensor,
\begin{equation} \label{energy-momentum-tensor}
	T^{\mu\nu} = (\epsilon + p)u^{\mu}u^{\nu} + pg^{\mu\nu} - 2\eta\sigma^{\mu\nu} - \zeta\Theta h^{\mu\nu} + u^{(\mu}q^{\nu)},
\end{equation}
that is expressed through thermodynamic forces and transport coefficients. The introduction of the transport coefficients provides us an opportunity to shed light on the different definitions of fluids that are possible in combination with the different values assumed by $\eta,\, \zeta,\, \kappa$. \cite{Rezzola}

\section{\label{sec:fieldeqs}Field equations, effective variables and the post-quasistatic approximation}
It is convenient to replace the energy density $\epsilon$ and isotropic pressure $p$ with effective variables
\begin{equation} \label{Effective-variables-eq}
	\begin{array}{l}
		\displaystyle \bar{\epsilon} = \epsilon + T_{\text{NPF}}{}^{0}_{\;0} \\[10pt]
		\displaystyle \bar{p} = p - T_{\text{NPF}}{}^{1}_{\;1}
	\end{array}
\end{equation}
so that Einstein's field equations 
\begin{equation} \label{Einstein-eq0}
	\displaystyle G^{\nu}_{\;\mu} = 8\pi T^{\nu}_{\;\mu},
\end{equation}
upon satisfying the metric \eqref{line-element}, provide five partial differential equations for each of the non-vanishing mixed-variance components of the stress--energy tensor, of which the four distinct ones are
\begin{subequations} \label{Einstein-eq}
	\begin{eqnarray}
		\!\!\!\!\!\!\!\! & 8\pi\bar{\epsilon} = \displaystyle e^{-\lambda}\left(\frac{\lambda'}{r} - \frac{1}{r^{2}}\right) + \frac{1}{r^{2}} \label{Einstein-eq1}
		\\
		\!\!\!\!\!\!\!\! & 8\pi\bar{p} = \displaystyle e^{-\lambda}\left(\frac{\nu'}{r}+ \frac{1}{r^{2}}\right) - \frac{1}{r^{2}} \label{Einstein-eq2}
		\\
		\!\!\!\!\!\!\!\! & 8\pi\tilde{p} = \displaystyle e^{-\lambda}\left(\frac{\nu''}{2} - \frac{\lambda'\nu'}{4} + \frac{\nu'^{2}}{4} + \frac{\nu' - \lambda'}{2r}\right) + \frac{e^{-\nu}}{4}\left(2\ddot{\lambda} + \dot{\lambda}(\dot{\lambda}-\dot{\nu})\right) \label{Einstein-eq3}
		\\
		\!\!\!\!\!\!\!\! & 8\pi T^{1}_{\;0} = \displaystyle e^{-\lambda}\dot{\lambda}/r, \label{Einstein-eq4}
	\end{eqnarray}
\end{subequations}
where $\tilde{p} = \bar{p} - T_{\text{NPF}}{}^{1}_{\;1} + T_{\text{NPF}}{}^{2}_{\;2}$ and overdots and primes denote partial differentiation with respect to $t$ and $r$, respectively. Another equation is proportional to \eqref{Einstein-eq3}, thus it is needless to consider it separately. The effective variables $(\bar{\rho},\, \bar{p})$ satisfy the same Einstein's field equations in the quasi-static regime as the corresponding physical variables $(\rho,\, p)$ (take account of the contribution of $T^{\mu\nu}_{\text{NPF}}$ to $p$ and $\epsilon$ in eq. \eqref{energy-density}). Therefore, the effective and physical variables share the same radial dependence. \cite{Herrera} Owing to the fact that \eqref{Einstein-eq1} involves only $\lambda$ and $\rho$, it becomes 
\begin{equation} \label{mass-eq}
	m' = 4\pi r^{2}\rho,
\end{equation}
once the new radial-dependent variable $m(r) = rt(1-e^{-\lambda})/2$ had been introduced. Similarly,
\begin{equation} \label{nu-equation}
	\nu = \nu_{0} + \int_{0}^{R}\frac{2(4\pi r^3 p + m)}{r(r-2m)}dr.
\end{equation}
Suppose that the radius extends to $R$, from \eqref{mass-eq} it is evident that the integral of the effective rest-mass density over the stellar interior
\begin{equation}
	m(R) = 4\pi\int_{0}^{R}\rho(r)r^{2}dr
\end{equation}
can be interpreted as the 'gravitational mass' of the system which includes all contributions to the relativistic mass (rest mass, internal energy, and the negative gravitational binding energy). However, integrating the total energy-density over the proper volume
\begin{equation} \label{volume}
	dV_{0} = \sqrt{\det(\gamma_{ij})}dx^{3} = 4\pi e^{\lambda/2}r^{2}dr,
\end{equation}
where the curvature of 3-space has been taken account of through $\lambda$, one obtains the mass
\begin{equation}
	\bar{m}(R) = 4\pi\int_{0}^{R}\rho(r)e^{\lambda/2}r^{2}dr
\end{equation}
that represents the sum of rest mass and internal energy. The difference between the two arises as a result of the mutual attraction of the fluid elements, called the binding energy, which is given by
\begin{equation}
	E_{B} = \bar{m}(R) - m(R) > 0
\end{equation}
and exhibits the amount of energy required to disassemble a whole system into separate elements.

It is convenient not to use eq. \eqref{Einstein-eq3} directly, but instead substracting it from \eqref{Einstein-eq2}. It can be replaced by the first-order ODE
\begin{equation}  
	\begin{array}{l}
		\bar{p}' + \frac{1}{2}(\bar{\rho} + \bar{p})\nu' = \displaystyle \frac{2(T_{\text{NPF}}{}^{1}_{\;1} - T_{\text{NPF}}{}^{2}_{\;2})}{r} + \frac{e^{-\nu}}{16\pi}\left(2\ddot{\lambda} + \dot{\lambda}(\dot{\lambda}-\dot{\nu})\right)
	\end{array}
\end{equation}
which stands for the condition for hydrostatic equilibrium, provided that the right-hand side of the equation is zero (cf. \eqref{pre-pulsation-eqs}). The first term on the right-hand side recovered from eq. \eqref{full-energy-momentum} is simply the result of
\begin{equation}
(g_{11}T^{11}_{\text{NPF}} - g_{22}T^{22}_{\text{NPF}}) = \frac{1}{4}\eta e^{-\nu/2}r\dot{\lambda}
\end{equation}
recovered from eq. \eqref{full-energy-momentum}. After the elimination of $\nu'$, one obtains the generalized Tolman--Oppenheimer--Volkoff (TOV) equation
\begin{equation} \label{modified-TOV}
	\begin{array}{l}
		\displaystyle \bar{p}' + \frac{(\bar{\rho} + \bar{p})(4\pi r^3 \bar{p} + m)}{r(r-2m)} = \frac{e^{-\nu}}{16\pi}\left(2\ddot{\lambda} + \dot{\lambda}(\dot{\lambda}-\dot{\nu} + 8\pi e^{\nu/2}\eta)\right)
	\end{array}
\end{equation}
for non-perfect fluids. Let us note that the non-zero terms on the left-hand side of eq. \eqref{modified-TOV} are not contradictory to conservation of energy and momentum in (\ref{conservation-of-mass}--\ref{conservation-of-energy}). It will be shown that a set of two equations \eqref{pre-pulsation-eqs} takes over its role; one describing the hydrostatic equilibrium and another describing the perturbation-induced departure from the equilibrium configuration. In order to close the system of the field equations \eqref{Einstein-eq} and the TOV eq. \eqref{modified-TOV}, it must be supplemented with an EOS relating some fundamental thermodynamical quantities. In general, the EOS takes the form
\begin{equation} \label{EOS-eq}
\bar{p} = \bar{p}(\bar{\epsilon},\, \bar{\rho}),
\end{equation}
where $\bar{\epsilon}$ and $\bar{p}$ are the total-energy density and isotropic pressure, given by eq. \eqref{Effective-variables-eq}.

\section{\label{sec:oscillations}Infinitesimal radial oscillations}
Let us suppose an equilibrium configuration of non-perfect fluid governed by the eqs. \eqref{Einstein-eq} is subject to a small linear perturbation that does not violate its spherical symmetry. Let $\delta$ hereafter denote a small ratio between the scale of variation of the perturbed variables and the correspondig ones in unperturbed configuration. Any quantity associated with the unperturbed equilibrium state is denoted by the subscript ''0'' whereas those that represent perturbations are equipped with the subscript ''1''. In respect to such a perturbation, motions in the radial directions arise. While formulating the equations governing the perturbed state, we shall ignore all quantities which are of the second or higher orders in motions. Consequently, the four-velocity of a fluid element will be expressed as
\begin{equation} \label{four-velocity}
	u_{\mu} = (-e^{\nu_{0}/2},\, e^{\lambda_{0}-\nu_{0}/2}\delta v_{1},\, 0,\, 0),
\end{equation}
where
\begin{equation} \label{velocity-perturbation}
	\delta v_{1} = dr/dt
\end{equation}
is the radial velocity with respect to the time coordinate $t$. 

\subsection{Perturbation equations for stellar oscillations}
One way to describe perturbations is the ''microscopic'' point of view where the observer follows individual fluid particles as they move through space and time. \cite{Friedman} It is convenient to introduce a displacement field $\xi_{1}$ (in our case, a scalar field) in the Lagrangian representation defined by
\begin{equation} \label{xi-definition}
	\frac{\partial\xi_{1}}{\partial t} = v_{1} + \frac{(T^{\text{NPF}}_{1})^{1}_{\;0}}{\bar{p}_{0} + \bar{\epsilon}_{0}},
\end{equation}
which connects fluid elements in the equilibrium with corresponding ones in the perturbed configration. There is yet another, ''macroscopic'' way of looking at the perturbations. \cite{Rampf} In the Eulerian representation of fluid motion, we simply consider changes in the variables $(\lambda,\, \nu,\, \epsilon,\, p)$ at a fluid element fixed in space and time. \cite{Constantin} This means that
\begin{equation} \label{lin-variables}
	\begin{array}{lll}
		\lambda = \lambda_{0} + \delta\lambda_{1}, \quad \nu = \nu_{0} + \delta\nu_{1}, \quad \epsilon = \epsilon_{0} + \delta\epsilon_{1}, \quad p = p_{0} + \delta p_{1},
	\end{array}
\end{equation}
where the linear perturbations $(\delta\lambda_{1},\, \delta\nu_{1}, \delta\epsilon_{1},\, \delta p_{1})$ are Euler changes. In this manner, retaining only those terms of the full stress--energy tensor \eqref{energy-momentum-tensor} that do not involve second or higher orders in $\delta$. The PF and NPF parts of the tensor assume the form of
	\begin{subequations} \label{energy-momentum-tensor-matrices}
		\begin{eqnarray}
			\hspace{-5em} (T_{\text{PF}})^{\mu}_{\;\nu} &\!\!\!\! = & \!\!\!\! 
			\begin{pmatrix}
				\epsilon_{0} + \delta\epsilon_{1} & (p_{0} + \epsilon_{0})\delta v_{1} & 0 & 0 \\
				-e^{\lambda_{0}-\nu_{0}}(p_{0} + \epsilon_{0})\delta v_{1} & -p_{0} - \delta p_{1} & 0 & 0 \\
				0 & 0 & -p_{0} - \delta p_{1} & 0  \\
				0 & 0 & 0 & -p_{0} - \delta p_{1}
			\end{pmatrix} \label{energy-momentum-tensor-matrices1}
			\\[5pt]
			\hspace{-5em} (T_{\text{NPF}})^{\mu}_{\;\nu} &\!\!\!\! = & \!\!\!\! 
			\begin{pmatrix}
				\delta (T^{\text{NPF}}_{1})^{0}_{\;0} & (T^{\text{NPF}}_{0})^{1}_{\;0} + \delta (T^{\text{NPF}}_{1})^{1}_{\;0} & 0 & 0 \\
				(T^{\text{NPF}}_{0})^{0}_{\;1} + \delta (T^{\text{NPF}}_{1})^{0}_{\;1} & \delta (T^{\text{NPF}}_{1})^{1}_{\;1} & 0 & 0 \\
				0 & 0 & \delta (T^{\text{NPF}}_{1})^{2}_{\;2} & 0  \\
				0 & 0 & 0 & \delta (T^{\text{NPF}}_{1})^{3}_{\;3}
			\end{pmatrix}, \label{energy-momentum-tensor-matrices2}
		\end{eqnarray}
	\end{subequations}
respectively. Notice that in eq. \eqref{energy-momentum-tensor-matrices1}, the off-diagonal elements are the only elements where the leading-order terms are absent, whereas in eq. \eqref{energy-momentum-tensor-matrices2} the leading-order terms are present only in these very same elements (see \hyperref[sec:stress-energy-components]{Appendix}). In view of the above, field equations \eqref{Einstein-eq1} and \eqref{Einstein-eq2} hold true if the pair of static metric potentials $(\lambda_{0},\, \nu_{0})$ is replaced by $(\lambda,\, \nu)$.
The corresponding linearized equations governing the perturbations are, thus,
\begin{equation} \label{Perturbed-Einstein-eq1}
	\frac{\partial}{\partial r}\left(r e^{-\lambda_{0}}\delta\lambda_{1}\right) = -8\pi r^{2}\delta\bar{\epsilon}_{1}
\end{equation}
and
\begin{equation} \label{Perturbed-Einstein-eq2}
	\frac{\partial}{\partial r}\delta\nu_{1} + \frac{\mathrm{d}\nu_{0}}{\mathrm{d}r}\delta\lambda_{1} = \delta\lambda_{1} - 8\pi e^{\lambda_{0}}r^{3}\delta\bar{p}_{1}.
\end{equation}
The appropriately linearized form of eqs. \eqref{Einstein-eq4} and \eqref{conservation-of-energy}, respectively,
\begin{equation} \label{Lin-Einstein-eq4}
	\frac{\partial}{\partial t}\delta\lambda_{1} = 8\pi e^{\lambda_{0}}r\left[(p_{0} + \epsilon_{0})\delta v_{1} + \delta (T^{\text{NPF}}_{1})^{1}_{\;0}\right]
\end{equation}
and
\begin{equation} \label{pre-pulsation-eqs0}
	\begin{array}{ll}
		& p'_{0} + \frac{1}{2}(p_{0} + \epsilon_{0})\nu'_{0} + e^{\lambda_{0}-\nu_{0}}(p_{0} + \epsilon_{0})\delta \dot{v}_{1} + (\delta p_{1} - \delta(T^{\text{NPF}}_{1})^{1}_{\;1})' + \frac{1}{2}(p_{0} + \epsilon_{0})\delta\nu'_{1} \\[5pt]
		& + \frac{1}{2}\left(\delta p_{1} + \delta\epsilon_{1} + \delta(T^{\text{NPF}}_{1})^{0}_{\;0} - \delta(T^{\text{NPF}}_{1})^{1}_{\;1}\right)\nu'_{0} - \frac{1}{2}(T^{\text{NPF}}_{0})^{0}_{\;1}(\delta\dot{\lambda}_{1} + \delta\dot{\nu}_{1}) \\[5pt]
		&  - \delta(\dot{T}^{\text{NPF}}_{1})^{0}_{\;1} + (\delta(T^{\text{NPF}}_{1})^{2}_{\;2}+\delta(T^{\text{NPF}}_{1})^{3}_{\;3}-2\delta(T^{\text{NPF}}_{1})^{1}_{\;1})/r = 0
	\end{array}
\end{equation}
shall serve in place of the remaining two field equations. At this point we may recall the definitions \eqref{Effective-variables-eq} for the effective variables $(\bar{\epsilon},\,\bar{p})$ and introduce their respective linearized forms $(\bar{\epsilon}_{0} + \delta\bar{\epsilon}_{1},\,\bar{p}_{0} + \delta\bar{p}_{1})$. 
Eqs. (\ref{lin-variables}--\ref{energy-momentum-tensor-matrices}) enable us to identify these new variables easily:
\begin{equation}
	\begin{array}{lll}
		\bar{\epsilon}_{0} &\!\!\!\! = & \!\!\!\! \epsilon_{0}, \quad \bar{\epsilon}_{1} = \epsilon_{1} + (T^{\text{NPF}}_{1})^{0}_{\;0}
		\\ 
		\bar{p}_{0} &\!\!\!\! = & \!\!\!\! p_{0}, \quad \bar{p}_{1} = p_{1} - (T^{\text{NPF}}_{1})^{1}_{\;1}.
	\end{array}
\end{equation}
Consequently, expressed in these more suitable variables, eq. \eqref{pre-pulsation-eqs0} decomposes into a set of two equations
\begin{equation} \label{pre-pulsation-eqs}
	\!\!\!\!\!\!\!\!\!\!\!\!\!\! \bar{p}'_{0} + \frac{\bar{p}_{0} + \bar{\epsilon}_{0}}{2}\nu'_{0} = 0, \quad 
	(\bar{p}_{0} + \bar{\epsilon}_{0})\left(e^{\lambda_{0}-\nu_{0}}\dot{v}_{1}  + \frac{\nu'_{1}}{2}\right) + \bar{p}'_{1} + \frac{\bar{p}_{1} + \bar{\epsilon}_{1}}{2}\nu'_{0} = \mathcal{T},
\end{equation}
where the first equation assumes the role of the constraint \eqref{modified-TOV} for hydrostatic equilibrium while the second one governs the dynamics of perturbation-induced departure from the equilibrium configuration. As will be shown later on in sec. \ref{sec-pulsation}, fluctuations of the stellar radius exhibit oscillatory behaviour: expansions and contractions in the outer layers as a star pursues to maintain equilibrium. The remaining NPF contributions are gathered on the right-hand side to form the source term
\begin{equation} \label{source-term1}
	\begin{array}{l}
		\mathcal{T} = \displaystyle \frac{1}{2}(T^{\text{NPF}}_{0})^{0}_{\;1}(\dot{\lambda}_{1} + \dot{\nu}_{1}) + (\dot{T}^{\text{NPF}}_{1})^{0}_{\;1} - \frac{(T^{\text{NPF}}_{1})^{2}_{\;2}+(T^{\text{NPF}}_{1})^{3}_{\;3}-2(T^{\text{NPF}}_{1})^{1}_{\;1}}{r},
	\end{array}
\end{equation}
which stems from frictional forces in the fluid itself. As such, it is responsible for exponential growth or damping, depending on the friction coefficient.

Now, in the sense of the definition \eqref{xi-definition}, the integration of eq. \eqref{Lin-Einstein-eq4} yields 
\begin{equation} \label{delta-lambd}
	\frac{e^{-\lambda_{0}}}{r}\delta\lambda_{1} = 8\pi(\bar{p}_{0} + \bar{\epsilon}_{0})\delta\xi_{1}
\end{equation}
or in consideration of the first eq. of \eqref{pre-pulsation-eqs},
\begin{equation} \label{delta-lambda}
	\delta\lambda_{1} = \delta\xi_{1}\frac{\mathrm{d}}{\mathrm{d}r}(\lambda_{0} + \nu_{0}).
\end{equation}
Eqs. \eqref{Perturbed-Einstein-eq1} and \eqref{delta-lambd} provide 
\begin{equation} \label{delta-epsilon}
	\delta\bar{\epsilon}_{1} = -\delta\xi_{1}\frac{\mathrm{d}\bar{\epsilon}_{0}}{\mathrm{d}r} - \delta\xi_{1}\frac{\mathrm{d}\bar{p}_{0}}{\mathrm{d}r} - \frac{1}{r^2}\frac{\partial(r^2\delta\xi_{1})}{\partial r}(\bar{p}_{0} + \bar{\epsilon}_{0}).
\end{equation}
Substituting for $d\bar{p}_{0}/dr$ in the last equation from the first eq. of \eqref{pre-pulsation-eqs}, we can also write
\begin{equation}
	\delta\bar{\epsilon}_{1} = -\delta\xi_{1}\frac{\mathrm{d}\bar{\epsilon}_{0}}{\mathrm{d}r} - \frac{e^{\lambda_{0}/2}}{r^2}\frac{\partial(r^2 e^{-\lambda_{0}/2}\delta\xi_{1})}{\partial r}(\bar{p}_{0} + \bar{\epsilon}_{0}).
\end{equation}
Considering next equation \eqref{Perturbed-Einstein-eq2} and substituting for $\delta\lambda_{1}$ in accordance with eq. \eqref{delta-lambd}, we obtain
\begin{equation}
	\frac{e^{-\lambda_{0}}}{r}\frac{\partial}{\partial r}\delta\nu_{1} = 8\pi\left[\delta\bar{p}_{1} + (\bar{p}_{0} + \bar{\epsilon}_{0})\left(\frac{\mathrm{d}\nu_{0}}{\mathrm{d}r} + \frac{1}{r}\right)\delta\xi_{1}\right]
\end{equation}
or in view of the first eq. of \eqref{pre-pulsation-eqs},
\begin{equation} \label{delta-nu}
	(\bar{p}_{0} + \bar{\epsilon}_{0})\frac{\partial}{\partial r}\delta\nu_{1} = \left[\delta\bar{p}_{1} + (\bar{p}_{0} + \bar{\epsilon}_{0})\left(\frac{\mathrm{d}\nu_{0}}{\mathrm{d}r} + \frac{1}{r}\right)\delta\xi_{1}\right]\frac{\mathrm{d}}{\mathrm{d}r}(\lambda_{0} + \nu_{0}).
\end{equation}

\subsection{Damping of stellar oscillations}
Suppose that for normal modes of the fluid perturbations $(\delta\lambda_{1},\, \delta\nu_{1},\, \delta\bar{\epsilon}_{1},\, \delta\bar{p}_{1})$ possess a harmonic time-dependence of the form $\exp(i\Omega t)$ where
\begin{equation} \label{complex-freq}
	i\Omega = -1/\tau + i\omega_{d}
\end{equation}
is a complex characteristic frequency to be determined in sec. \ref{sec-pulsation}. Being subject to the damping effect of the dissipative forces \eqref{constitutive-eqs}, the fluid perturbations oscillate with a damped angular frequency (sometimes called pseudo-frequency)
\begin{equation} \label{damping-freq}
	\omega_{d} = \omega_{n}\sqrt{1 - \zeta^2},
\end{equation}
related to $\omega_{n}$ which is the natural frequency (or resonant frequency) of the undamped system. The rate at which the normal modes of radial oscillations are damped is characterized by the relaxation time (or damping time)
\begin{equation} \label{decay-rate}
	\tau = -1/\omega_{n}\zeta
\end{equation}
and can be determined from energy-dissipation eq. \eqref{energy-dissipation}. \cite{Lindblom1983} Expressed by the damping ratio $\zeta$, a dimensionless non-negative parameter, the complex characteristic frequency is
\begin{equation} \label{complex-freq2}
	i\Omega = -\omega_{n}(\zeta - i\sqrt{1 - \zeta^2}).
\end{equation}
The value of $\zeta$ prescribes the frequency response and critically determines the dynamical behaviour. The oscillation is undamped $(\zeta = 0)$, if the star oscillates with its natural angular frequency $\omega_{n}$. It is underdamped $(\zeta < 1)$, if the star oscillates with a damped frequency $\omega_{d}$ and with the amplitude gradually decreasing with the rate of decay $1/\tau$.	It is critically damped $(\zeta = 1)$, if the star returns to steady state as quickly as possible without any oscillation.

With the harmonic time-dependence, we can rewrite the second equation of \eqref{pre-pulsation-eqs} in the form
\begin{equation} \label{pre-pulsation-eqs3}
\begin{array}{l}
\displaystyle \Omega^{2}e^{\lambda_{0}-\nu_{0}}(\bar{p}_{0} + \bar{\epsilon}_{0})\xi_{1} = \bar{p}'_{1} + \left(\frac{\lambda'_{0}}{2} + \nu'_{0}\right) + \frac{\bar{\epsilon}_{1}\nu'_{0}}{2} - \frac{\bar{p}_{0} + \bar{\epsilon}_{0}}{2}\left(\nu'_{0} + \frac{1}{r}\right)(\lambda'_{0} + \nu'_{0})\xi_{1} + i\Omega\mathcal{S}_{1},
\end{array}
\end{equation}
where we have substituted for $(\bar{p}_{0} + \bar{\epsilon}_{0})\nu'_{1}$ in accordance with eq. \eqref{delta-nu}. Also, it should be recalled that $\bar{\epsilon}_{1}$ is expressed in terms of $\xi_{1}$ and the perturbed variables by eq. \eqref{delta-epsilon}. The additional term $\mathcal{S}_{1}$ is given by the expression \eqref{appendix-source-term2}.

\section{\label{sec:gamma} The conservation of baryon number}
The continuity equation \eqref{conservation-of-mass} of the rest-mass density current given as in \eqref{perfect-energy-momentum-tensor} involves the conservation of the baryon number, represented by the condition
\begin{equation}
	\nabla_{\mu}(Nu^{\mu}) = 0,
\end{equation}
provided that $N$ is the number of constituent baryons per unit volume. Let the covariant derivative be written differently, in the form
\begin{equation} \label{conservation-of-baryon-number}
	\frac{\partial}{\partial x^{\mu}}(Nu^{\mu}) + (Nu^{\mu})\frac{\partial}{\partial x^{\mu}}\log\sqrt{-g} = 0
\end{equation}
with the expression $\partial(\log\sqrt{-g})/\partial x^{\mu}$ being put in the place of the Christoffel symbols corresponding to the metric given in eq. \eqref{line-element} and
\begin{equation}
	g = e^{\lambda + \nu}r^4\sin^2\theta
\end{equation}
being the determinant of the metric tensor. In the framework of the present linearized theory, it is quite reasonable to define the baryon number by
\begin{equation}
	N = N_{0} + \delta N_{1},
\end{equation}
so that eq. \eqref{conservation-of-baryon-number} endowed with the non-vanishing components of the four-velocity \eqref{four-velocity} emerges as
\begin{equation}
	\displaystyle e^{-\nu_{0}/2}\frac{\partial}{\partial t}\delta N_{1} + \frac{1}{r^2}\frac{\partial}{\partial r}\left(N_{0}r^2e^{-\nu_{0}/2}\delta v_{1}\right) + \frac{1}{2}N_{0}e^{-\nu_{0}/2}\left(\frac{\partial}{\partial t}\delta\lambda_{1} + \delta v_{1}\frac{\mathrm{d}}{\mathrm{d}r}(\lambda_{0}+\nu_{0})\right) = 0.
\end{equation}
With $v_{1}$ replaced by the Lagrangian displacement $\xi_{1}$ defined in eq. \eqref{xi-definition}, the last equation brings about fluctuations in the baryon number:
\begin{equation} \label{conservation-of-baryon-number3}
	\displaystyle\delta N_{1} + \frac{e^{\nu_{0}/2}}{r^2}\frac{\partial}{\partial r}\left(N_{0}r^2e^{-\nu_{0}/2}\delta\xi_{1}\right) + \frac{1}{2}N_{0}\left(\delta\lambda_{1} + \delta\xi_{1}\frac{\mathrm{d}}{\mathrm{d}r}(\lambda_{0}+\nu_{0})\right) = i\Omega^{-1}\mathcal{N},
\end{equation}
where the second term on the right-hand side vanishes on account of eq. \eqref{delta-lambda} and
\begin{equation}
	\mathcal{N} = \frac{1}{r^2}e^{\lambda_{0}+2\nu_{0}}\frac{\partial}{\partial r}\left(r^2e^{-(\lambda_{0}+2\nu_{0})/2}N_{0}\frac{(T^{\text{NPF}}_{1})^{1}_{\;0}}{\bar{p}_{0} + \bar{\epsilon}_{0}}\right)
\end{equation}
represents the effect of dissipative terms on the baryon-number perturbation. Subsequently, eq. \eqref{conservation-of-baryon-number3} reduced significantly and one obtains
\begin{equation} \label{delta-n}
	\delta N_{1} = -\frac{\mathrm{d}N_{0}}{\mathrm{d}r}\delta\xi_{1} - N_{0}\frac{e^{\nu_{0}}}{r^2}\frac{\partial}{\partial r}\left(r^2e^{-\nu_{0}}\delta\xi_{1}\right) - i\Omega^{-1}\mathcal{N}.
\end{equation}
Provided that $N \equiv N(\epsilon,\, p)$ is an EOS that corresponds to \eqref{EOS-eq}, small linear perturbations in the energy density or pressure treated as variation are expected to induce baryon-number perturbations given by
\begin{equation}
	\delta \bar{N}_{1} = \frac{\partial \bar{N}_{1}}{\partial\bar{\epsilon}_{1}}\delta\bar{\epsilon}_{1} + \frac{\partial \bar{N}_{1}}{\partial \bar{p}_{1}}\delta \bar{p}_{1},
\end{equation}
which in turn yields
\begin{equation} \label{delta-p1}
	\delta \bar{p}_{1} = \left(\frac{\partial \bar{N}_{0}}{\partial \bar{p}_{0}}\right)^{-1}\left(\delta \bar{N}_{1} - \frac{\partial \bar{N}_{0}}{\partial \bar{\epsilon}_{0}}\delta\bar{\epsilon}_{1}\right),
\end{equation}
under the assumption that variables in the perturbed state relate to each other roughly the same way as the corresponding variables in equilibrium, that is $\partial \bar{N}_{1}/\partial\bar{\epsilon}_{1} \approx \partial \bar{N}_{0}/\partial\bar{\epsilon}_{0}$ and $\partial \bar{N}_{1}/\partial \bar{p}_{1} \approx \partial \bar{N}_{0}/\partial \bar{p}_{0}$. With $\delta\bar{\epsilon}_{1}$ and $\delta \bar{N}_{1}$ given by eqs. \eqref{delta-epsilon} and \eqref{delta-n}, respectively, eq. \eqref{delta-p1} comes to be
\begin{equation} \label{delta-p2}
	\delta \bar{p}_{1} = -\frac{\mathrm{d}\bar{p}_{0}}{\mathrm{d}r}\delta\xi_{1} - \Gamma\bar{p}_{0}\frac{e^{\nu_{0}}}{r^2}\frac{\partial}{\partial r}\left(r^2 e^{-\nu_{0}/2}\delta\xi_{1}\right) - i\Omega^{-1}\left(\frac{\partial \bar{N}_{0}}{\partial \bar{p}_{0}}\right)^{-1}\mathcal{S}_{2},
\end{equation}
where $\Gamma$ is identified with the adiabatic index given by
\begin{equation} \label{adiabatic-index}
	\Gamma = \left(\bar{p}_{0}\frac{\partial \bar{N}_{0}}{\partial r}\right)^{-1}\left(\bar{N}_{0} - (\bar{p}_{0}+\bar{\epsilon}_{0})\frac{\partial \bar{N}_{0}}{\partial\bar{\epsilon}_{0}}\right).
\end{equation}
Within a star that is dynamically stable against infinitesimal radial adiabatic perturbations, the average value for the adiabatic index is greater than or equal to $4/3$. Expressly,
\begin{equation}
	\Gamma = \frac{\epsilon+p}{p}c_{s}^{2} \geq \frac{4}{3},
\end{equation}
where $\epsilon$ denotes the  total-energy density, $p$ the radial pressure, and $c_{s}$ the radial sound speed, respectively \cite{Ivanov2017}.

\section{The pulsation equation and eigenvalue problem} \label{sec-pulsation}
With elimination of $\bar{\epsilon}_{1}$ and $\bar{p}_{1}$ from eq. \eqref{pre-pulsation-eqs3}, through eqs. \eqref{delta-epsilon} and \eqref{delta-p2}, it emerges as
\begin{equation} \label{pre-pulsation-eqs4}
\begin{array}{l}
	\!\!\! \displaystyle \Omega^{2}e^{\lambda_{0}-\nu_{0}}(\bar{p}_{0} + \bar{\epsilon}_{0})\xi_{1} = \frac{\mathrm{d}}{\mathrm{d}r}(\xi_{1}\bar{p}'_{0}) - \left(\frac{1}{2}\lambda'_{0} + \nu'_{0}\right)\xi_{1}\bar{p}'_{0} - \frac{1}{2}(\bar{p}_{0} + \bar{\epsilon}_{0})\left(\nu'_{0} + \frac{1}{r}\right)(\lambda'_{0} + \nu'_{0}) \\[10pt]
	\!\!\! \displaystyle + \frac{\nu'_{0}}{2 r^{2}}\frac{\mathrm{d}}{\mathrm{d}r}\left[r^{2}(\bar{p}_{0} + \bar{\epsilon}_{0})\xi_{1}\right] - e^{-(\lambda_{0}+2\nu_{0})/2}\frac{\mathrm{d}}{\mathrm{d}r}\left[e^{(\lambda_{0}+3\nu_{0})/2}\frac{\Gamma\bar{p}_{0}}{r^2}\frac{\mathrm{d}}{\mathrm{d}r}\left(r^2 e^{-\nu_{0}/2}\xi_{1}\right)\right] + i(\Omega \mathcal{S}_{1} - \Omega^{-1}\mathcal{S}_{2}),
\end{array}
\end{equation}
where the third and forth terms on the right-hand side are the result of the transformation	$\frac{\mathrm{d}f}{\mathrm{d}r} + f\frac{\mathrm{d}g}{\mathrm{d}r} = \exp(-g)\frac{\mathrm{d}}{\mathrm{d}r}\left[f\exp(g)\right]$, which applies to any two functions $(f,\, g)$ of the variable $r$ and $\mathcal{S}_{2}$ is written out in eq. \eqref{S2-def} in appendix. 

Substituting for $\bar{p}'_{0}$ from the first eq. of \eqref{pre-pulsation-eqs} and applying the same transformation for the first two terms on the right-hand side of eq. \eqref{pre-pulsation-eqs4} allow us to merge them with the next two terms in the expression
\begin{equation}
	\frac{1}{2}(\bar{p}_{0} + \bar{\epsilon}_{0})\left(\nu''_{0} - \frac{1}{2}\lambda'_{0}\nu'_{0} - \frac{1}{2}\lambda'_{0} - \frac{3}{r}\nu'_{0}\right)\xi_{1},
\end{equation}
which (cf. (56) in \cite{Chandrasekhar}), in turn, compared with the field equation \eqref{Einstein-eq3} restricted by equilibrium conditions,
\begin{equation}
	8\pi \bar{p}_{0} = \displaystyle e^{-\lambda}\left(\frac{\nu''_{0}}{2} - \frac{\lambda'_{0}\nu'_{0}}{4} + \frac{\nu'_{0}{}^{2}}{4} + \frac{\nu'_{0} - \lambda'_{0}}{2r}\right),
\end{equation}
is reduced to three terms. Making use of the first eq. of \eqref{pre-pulsation-eqs} once again, the relation \eqref{pre-pulsation-eqs4} appreciably reduces to give the pulsation equation
\begin{equation} \label{pulsation-eq}
	\begin{array}{l}
		\Omega^{2}e^{\lambda_{0}-\nu_{0}}(\bar{p}_{0} + \bar{\epsilon}_{0})\xi_{1} = \displaystyle \left(\frac{4}{r}\frac{\mathrm{d}\bar{p}_{0}}{\mathrm{d}r} + 8\pi e^{\lambda_{0}}\bar{p}_{0}(\bar{p}_{0} + \bar{\epsilon}_{0}) - \frac{1}{\bar{p}_{0} + \bar{\epsilon}_{0}}\left[\frac{\mathrm{d}\bar{p}_{0}}{\mathrm{d}r}\right]^{2}\right)\xi_{1}
		\\[10pt]
		\displaystyle \quad\quad\quad - e^{-(\lambda_{0}+2\nu_{0})/2}\frac{\mathrm{d}}{\mathrm{d}r}\left(e^{(\lambda_{0}+3\nu_{0})/2}\frac{\Gamma \bar{p}_{0}}{r^{2}}\frac{\mathrm{d}}{\mathrm{d}r}\left[r^{2}e^{-\nu_{0}/2}\xi_{1}\right]\right) + i(\Omega \mathcal{S}_{1} - \Omega^{-1}\mathcal{S}_{2})
	\end{array}
\end{equation}
associated to the class of second-order linear ODEs. Besides this, the definitions (\ref{complex-freq}--\ref{complex-freq2}) make possible the separation of real and imaginary parts of the complex frequency squared:
\begin{equation} \label{omega1}
	\Omega^2 = \omega_{n}^2(1-2\zeta^2) + 2i\omega_{n}^2\zeta\sqrt{1-\zeta^2},
\end{equation}
thus the last term of the right-hand side:
\begin{equation} \label{omega2}
	i(\Omega \mathcal{S}_{1} - \Omega^{-1}\mathcal{S}_{2}) = \displaystyle -\zeta\left(\omega_{n}\mathcal{S}_{1} + \omega_{n}^{-1}\mathcal{S}_{2}\right) + i\sqrt{1-\zeta^2}\left(\omega_{n}\mathcal{S}_{1} - \omega_{n}^{-1}\mathcal{S}_{2}\right).
\end{equation}
For positive $\Omega^{2}$, the characteristic frequency $\Omega$ is real and thus, the solution is purely oscillatory. However, for $\Omega^2 < 0$, $\Omega$ contains an imaginary part, which corresponds to a damped solution. Since the general solution is a superposition of damped modes, the occurrence of a negative value of $\Omega^2$ corresponds to a secular instability whose growth time is long compared to the dynamical time of radial oscillations. The Harrison--Zel'dovich--Novikov criterion \cite{Harrison1965} for static stability of compact stars states that the total mass of such stars increases with the central density $\rho_{0}$ which implies that $dM(\rho_{0})/d\rho_{0} \geq 0$ for the stable region where $M(\rho_{0})$ is the function of total mass in terms of the central density. For neutron stars this will, indeed, happen for $\rho_{0}$ larger than the critical central density $\rho_{\text{crit}}$ at which the stellar mass $M(\rho_{0})$ as a function of $\rho_{0}$ has its maximum. In this case the star will ultimately collapse to a black hole. For $\rho_{0} = \rho_{\text{crit}}$ there must be a neutral mode with the corresponding eigenvalue $\omega^{2} = 0$. \cite{Ivanov2017}

\subsection{Eigenfrequencies of radial pulsation} \label{sec:eigenfrequency}
The real part of left-hand side of eq. \eqref{pulsation-eq} can be equated to the real part of right-hand side, comparably to the imaginary parts. Intrinsically, we can recast it in the so-called homogeneous Sturm--Liouville form
\begin{equation} \label{pulsation-eq2}
	\frac{\mathrm{d}}{\mathrm{d}r}\left[\mathcal{P}\frac{\mathrm{d}\chi}{\mathrm{d}r}\right] + \left[\mathcal{Q} + \Lambda_{n}\mathcal{R}\right]\chi = 0
\end{equation}
with a free parameter
\begin{equation} \label{eigen-value}
	\Lambda_{n} \equiv (1-2\zeta^2)\omega_{n}^2
\end{equation}
denoting the eigenvalues and a set of coefficient functions
\begin{equation} \label{operators}
\begin{split}
& \mathcal{P}(r) = \displaystyle r^{-2}e^{(\lambda_{0}+3\nu_{0})/2}\Gamma \bar{p}_{0}
\\
& \mathcal{Q}(r) = \displaystyle r^{-2}e^{(\lambda_{0}+3\nu_{0})/2}\bigg[\left(\frac{\bar{p}'_{0}}{\bar{p}_{0}+\bar{\epsilon}_{0}}-\frac{4}{r}\right)\bar{p}'_{0} + 8\pi e^{(\lambda_{0}+3\nu_{0})/2}\bar{p}_{0}(\bar{p}_{0}+\bar{\epsilon}_{0})\bigg]
\\
& \mathcal{R}(r) = r^{-2}e^{(3\lambda_{0}+\nu_{0})/2}(\bar{p}_{0} + \bar{\epsilon}_{0})
\end{split}
\end{equation}
hailing from eq. \eqref{pulsation-eq}, are specified at the outset. The function $\mathcal{R}(r)$ is referred to as weighting function. The normalized Lagrangian displacement defined by
\begin{equation} \label{chi-definition}
	\chi \equiv r^{2}e^{-\nu_{0}}\xi_{1}
\end{equation}
is a scalar-valued function of the variables $(t,\, r)$. Provided that $\chi$ satisfies eq. \eqref{pulsation-eq2}, it is called a solution. Solutions of \eqref{pulsation-eq2} are subject to the boundary conditions
\begin{equation} \label{boundary-condition}
	\chi = 0 \text{ at } r = 0 \quad \text{and} \quad \delta p = 0 \text{ at } r = R.
\end{equation}
Together with the boundary condition \eqref{boundary-condition}, the pulsation equation \eqref{pulsation-eq2} imposes a Sturm--Liouville eigenvalue problem (SLEVP), which seeks non-trivial solutions only for a countable set of real eigenvalues $\{\Lambda_{1}, \Lambda_{2}, \ldots, \Lambda_{n}\}$. The SLEVP is said to be regular if $\mathcal{P}> 0$ and $\mathcal{R} > 0$ for any $r \in [0,\, R]$, the functions $(\mathcal{P},\, \mathcal{P}',\, \mathcal{Q},\, \mathcal{R},\, \mathcal{S})$ are continuous over the finite interval $[0,\, R]$, and the problem has separated boundary conditions of the form
\begin{subequations} \label{boundary-condition2}
	\begin{eqnarray}
		\alpha_{1}\chi(0) + \alpha_{2}\chi'(0) & = & 0 \quad \text{ for } \quad \alpha_{1}^2 + \alpha_{2}^2 > 0
		\\
		\beta_{1}\chi(R) + \beta_{2}\chi'(R) & = & 0 \quad \text{ for } \quad \beta_{1}^2 + \beta_{2}^2 > 0.
	\end{eqnarray}
\end{subequations}
Sturm--Liouville theory states that the eigenvalues of the regular SLEVP are real and can be arranged in ascending order such that
\begin{equation} \label{eigenvalue-spectrum}
	\Lambda_{1} < \Lambda_{2} < \Lambda_{3} < \ldots < \Lambda_{n} < \ldots \quad \text{where} \quad \lim_{n \to \infty}\Lambda_{n} = +\infty.
\end{equation}
Corresponding to each eigenvalue $\Lambda_{n}$ is a unique (up to a normalization constant) eigenfunction $\chi_{n}(r)$ which has exactly $n - 1$ zeros in $(0,R)$. Moreover, the normalized eigenfunctions form an orthonormal basis
\begin{equation}
\int^{0}_{R}\chi_{n}(r)\chi_{m}(r)\mathcal{R}(r)dr = \delta_{mn},
\end{equation}
where $\delta_{mn}$ is the Kronecker delta and to each $\Lambda_{n}$ is associated with a single eigenfunction $\chi_{n}$.

\subsection{Characteristic relaxation time of radial pulsation}\label{sec:relaxation-time}
In the Newtonian limit, the kinetic energy contained in these oscillations is given by
\begin{equation} \label{kinetic-energy}
E_{k} = \frac{1}{2}\int\rho\delta v_{1} \delta v_{1}^{*} dV_{0},
\end{equation}
an integralover an element of proper volume of fluid $dV_{0}$ (see eq. \eqref{volume}), where $\delta v_{1}^{*}$ is the complex conjugate of the velocity perturbation $\delta v_{1}$. Associated with the radial displacement $\delta\xi_{1} = \delta r/r$, the later is given for radial oscillations by eq. \eqref{xi-definition} as 
\begin{equation} \label{velocity-perturbation2}
\delta v_{1} = i\Omega\left(\delta\xi_{1}  - \frac{\exp(-\nu_{0}/2)}{\bar{p}_{0} + \bar{\epsilon}_{0}}\left[3\eta\nu'_{0}\delta\nu_{1}-\left(\eta-\frac{1}{4}\kappa T\right)\delta\nu'_{1}\right]\right),
\end{equation}
where, in accordance with second expression of \eqref{t1components}, we have substituted for $(\delta T^{\text{NPF}}_{1})^{1}_{\;0}$. The total energy in an oscillating star consists of kinetic and potential energy which are supposed to equally contribute to the total energy of harmonic oscillations, thus given by $E = 2E_{k}$. On account of the density, $\rho$, is reasonably uniform in neutron stars \cite{Cutler}, the total energy contained in the oscillation is given as
\begin{equation}
E = \bar{\rho}\omega_{n}\epsilon^{2}R^{5}
\end{equation}
by evaluating the integral \eqref{kinetic-energy} explicitly for an avaraged density $\bar{\rho} = 3M/4\pi R^{3}$. Being bilinear in the fluid perturbations, $E$ has a time-dependence $\exp[-2\operatorname{Im}(\Omega)t]$. \cite{Lindblom1983} Subsequently, its time derivative implies  that 
\begin{equation}
\frac{\mathrm{d}E}{\mathrm{d}t} = -2\operatorname{Im}(\Omega)E,
\end{equation}
which together with the energy-dissipation rate for the stress--energy tensor \eqref{energy-momentum-tensor} as
\begin{equation} \label{energy-dissipation}
- \frac{\mathrm{d}E}{\mathrm{d}t} = -\int\left(2\eta\delta\sigma^{\mu\nu}\delta\sigma^{*}_{\mu\nu} + \zeta(\delta\Theta)^{2} + \frac{\kappa}{T}\nabla_{\mu}\delta T \nabla^{\mu} \delta T^{*}\right)dV_{0},
\end{equation}
directly determines the dissipative time-scale of small perturbations of the fluid away from the equilibrium state as
\begin{equation} \label{dissipative-time-scale}
\tau = -2E/\dot{E}.
\end{equation}
An approximate formula for each dissipative time-scales can be given by evaluating the corresponding dissipation integral \eqref{energy-dissipation}. Cutler \& Lindblom \cite{Cutler} have found that the following propotionalities hold for the dissipative time-scales:
\begin{equation} \label{dissipative-time-scale2}
\frac{1}{\tau_{\eta}} \sim \frac{\eta}{\rho R^{2}}, \quad \frac{1}{\tau_{\zeta}} \sim \frac{\eta}{\rho R^{2}}, \quad \frac{1}{\tau_{\kappa}} \sim \frac{\kappa T}{\rho^{2}R^{4}}
\end{equation}
which reveals that the time-scale of shear viscosity is much shorter than either that of bulk viscosity or thermal conductivity. The imaginary part of the pulsation equation \eqref{pulsation-eq}, together with eqs. (\ref{omega1}--\ref{omega2}) yields
\begin{equation} \label{zeta-eq}
\zeta_{n} = \frac{\mathcal{S}_{2} - \omega_{n}^{2}\mathcal{S}_{1}}{2\omega_{n}^{3}(\bar{p}_{0} + \bar{\epsilon}_{0})\xi_{1}}e^{\nu_{0}-\lambda_{0}}
\end{equation}
which implies that a unique damping ratio $\zeta_{n}$ corresponds to each eigenfrequency $\omega_{n}$ computed for undamped oscillations. From \eqref{zeta-eq}, it is evident that higher frequency components die out first. These damping ratios are identical with those associated with the dissipative time-scales \eqref{dissipative-time-scale2} which, according with \eqref{decay-rate}, are obtained from $\zeta_{n} = \tau_{n}\omega_{n}$.

\section{Concluding remarks} \label{sec:conclusion}
A generic formulation of the dynamical equations governing small adiabatic radial oscillations of pulsating relativistic stars has been proposed in this paper through a perturbation scheme that, combined with the equations of viscous thermally-conductive fluids, constitutes an extension of radially pulsating perfect-fluid stellar models.  We have proved that, similarly to the regular perfect-fluid case, the stellar pulsation equation \eqref{pulsation-eq} expressed by a set of effective variables \eqref{Effective-variables-eq} which involve dissipative terms, can be recast in a homogenous Sturm--Liouville form \eqref{pulsation-eq2} with separated boundary conditions \eqref{boundary-condition2}. In contrast to the regular perfect-fluid case, the associated eigenvalue problem is generalized for a discrete set of eigenfunctions with complex eigenvalues where the real and imaginary parts of the eigenvalues represent the squared natural frequency and relaxation time (or decay rate) of the oscillation, respectively. In the absence of dissipation, the discrete spectrum consists of real eigenvalues that form a complete set. 

The main novelty of this approach is the ability to directly relate the damping ratio of oscillations to the expressions $\mathcal{S}_{1}$ and $\mathcal{S}_{2}$ in (\ref{S1-def}--\ref{S2-def}), which stem from the viscous and heat-conductive contributions to the stress--energy tensor, without relying on explicit numerical computations. An illustrative example set by \cite{Cutler} for neutron star with uniform density, allowed us in this paper to estimate the rate at which the viscosity and thermal conductivity of the nuclear matter drains energy from the oscillations. In accordance with the literature, the time-scale of shear viscosity is much shorter than either that of bulk viscosity or thermal conductivity and the imaginary part of the pulsation equation indicate that higher components vanish first from the frequency spectrum. The usefulness of our analytical approximation method is evidently restricted to providing qualitative and ``order-of-magnitude'' information about the dissipative time-scales in \eqref{dissipative-time-scale2} rather than a precise one.

\appendix
\section*{Appendix} \label{sec:stress-energy-components}
\renewcommand{\theequation}{A.\arabic{equation}}

We are to enumerate the non-vanishing components of the NPF stress--energy tensor $ (T_{\text{NPF}})^{\mu}_{\;\nu}$ referred in eq. \eqref{energy-momentum-tensor-matrices2}. To explicitly evaluate them, the arithmetic operations in eqs. (\ref{shear-tensor}--\ref{energy-momentum-tensor}) have to performed. The non-vanishing components of the zeroth-order are
\begin{equation} \label{t0components}
	\displaystyle (T^{\text{NPF}}_{0})^{1}_{\;0} = -e^{\lambda_{0}-\nu_{0}}(T^{\text{NPF}}_{0})^{0}_{\;1} = \frac{1}{4}e^{-\nu_{0}/2}\left[8\kappa T'+\left(12\eta-\kappa T\right)\nu'_{0}\right],
\end{equation}
those of the first-order are
\begin{equation} \label{t1components}
	\begin{array}{l}
		\displaystyle \frac{(T^{\text{NPF}}_{1})^{2}_{\;2} + (T^{\text{NPF}}_{1})^{3}_{\;3} - 2(T^{\text{NPF}}_{1})^{1}_{\;1}}{r} = \frac{e^{-\nu_{0}/2}}{r}\left(A_{0}v_{1}+\eta\dot{\lambda}_{1}\right) \\[10pt]
		\displaystyle (T^{\text{NPF}}_{1})^{1}_{\;0} = -e^{\lambda_{0}-\nu_{0}}(T^{\text{NPF}}_{1})^{0}_{\;1} = e^{-\nu_{0}/2}\left[3\eta\nu'_{0}\nu_{1}-\left(\eta-\frac{1}{4}\kappa T\right)\nu'_{1}\right]
	\end{array}
\end{equation}
with an only radial-dependent coefficient
\begin{equation}
	A_{0} = \eta\left[\left(1+3e^{\nu_{0}-\lambda_{0}}\right)\nu'_{0} - 9\lambda'_{0} - \frac{2}{r}e^{\lambda_{0}}\right] + \kappa\left[T\nu'_{0}-8T'\right].
\end{equation}
In view of \eqref{t1components}, it is evident that to abide by the nature of harmonic time-dependence, a relation of the form
\begin{equation}
	(\dot{T}^{\text{NPF}}_{1})^{1}_{\;0} = i\frac{\Omega}{c}(T^{\text{NPF}}_{1})^{1}_{\;0}
\end{equation}
is implied. According to the definition \eqref{xi-definition}, this implication requires
\begin{equation} \label{v-dot-definition}
	\dot{v}_{1} = -\frac{\Omega^{2}}{c^{2}}\xi_{1} - i\frac{\Omega}{c}\frac{(T^{\text{NPF}}_{1})^{1}_{\;0}}{\bar{p}_{0} + \bar{\epsilon}_{0}}
\end{equation}
to hold. In regard to to eq. \eqref{v-dot-definition}, the expression $e^{\lambda_{0}-\nu_{0}}(\dot{T}^{\text{NPF}}_{1})^{1}_{\;0}$ is combined with $\mathcal{T}$ given by eq. \eqref{source-term1} to compose
\begin{equation} \label{appendix-source-term2}
	\mathcal{S}_{1} = -\frac{1}{2}(T^{\text{NPF}}_{0})^{0}_{\;1}(\dot{\lambda}_{1} + \dot{\nu}_{1}) + \displaystyle\frac{(T^{\text{NPF}}_{1})^{2}_{\;2}+(T^{\text{NPF}}_{1})^{3}_{\;3}-2(T^{\text{NPF}}_{1})^{1}_{\;1}}{r}
\end{equation}
which first appears in eq. \eqref{pre-pulsation-eqs3}. Substituted with the set of eqs. (\ref{t0components}--\ref{v-dot-definition}), the expression unfolds as
\begin{equation} \label{S1-def}
	\mathcal{S}_{1} = \displaystyle \frac{e^{-\nu_{0}/2}}{r}\left(\eta\lambda_{1} + A_{0}\xi_{1} - A_{0}\frac{(T^{\text{NPF}}_{1})^{1}_{\;0}}{\bar{p}_{0} + \bar{\epsilon}_{0}}\right) - \frac{1}{2}(T^{\text{NPF}}_{0})^{1}_{\;0}(\lambda_{1} + \nu_{1}).
\end{equation}
Next to the $\mathcal{S}_{1}$, another quantity, expressed by
\begin{equation} \label{S2-def}
\begin{array}{ll}
\mathcal{S}_{2} = & \!\!\displaystyle e^{-(\lambda_{0}+2\nu_{0})/2}\frac{\mathrm{d}}{\mathrm{d}r}\bigg[\frac{e^{\lambda_{0}+2\nu_{0}}}{r^2}\left(\frac{\partial \bar{N}_{0}}{\partial\bar{p}_{0}}\right)^{-1}\frac{\mathrm{d}}{\mathrm{d}r}\left(r^2 e^{-(\lambda_{0}+2\nu_{0})/2}\bar{N}_{0}\frac{(T^{\text{NPF}}_{1})^{1}_{\;0}}{\bar{p}_{0} + \bar{\epsilon}_{0}}\right)\bigg].
\end{array}
\end{equation}
appears in different forms the pulsation equation, namely in eqs. \eqref{pulsation-eq} and \eqref{pre-pulsation-eqs4}. Together, they constitute the expression $i(\Omega \mathcal{S}_{1} - \Omega^{-1}\mathcal{S}_{2})$ which represents the source of inhomogenity of the SL equation.

\section*{Acknowledgments}
I would like to gratefully acknowledge the financial support from ''PHAROS'', COST Action CA16214. I am also grateful for K. Kokkotas for valuable discussions and for his generous hospitality at the University of T\"{u}bingen in April 2018. \\

\bibliography{cikk4}

\end{document}